\documentclass[aps,pre,twocolumn,showpacs]{revtex4}
\usepackage{amsmath,graphicx,amsbsy,amssymb,amsmath,epsfig,latexsym,wasysym,pifont,mathrsfs}
\usepackage{float} 
\newfloat{widefig}{thp}{lop}
\usepackage{comment} 
\usepackage{verbatim}

\begin{document}
\title{Critical properties of a dissipative sandpile model on small
  world networks}

\author{Himangsu Bhaumik} 
\email{himangsu@iitg.ac.in}
\author{S. B. Santra}
\email{santra@iitg.ac.in}
\affiliation{Department of Physics, Indian Institute of Technology
  Guwahati, Guwahati-781039, Assam, India.}

\date{\today}
 
\begin{abstract}  A dissipative sandpile model (DSM) is constructed
  and studied on small world networks (SWN). SWNs are generated adding
  extra links between two arbitrary sites of a two dimensional square
  lattice with different shortcut densities $\phi$. Three different
  regimes are identified as regular lattice (RL) for $\phi\lesssim
  2^{-12}$, SWN for $2^{-12}<\phi< 0.1$ and random network (RN) for
  $\phi\ge 0.1$. In the RL regime, the sandpile dynamics is
  characterized by usual Bak, Tang, Weisenfeld (BTW) type correlated
  scaling whereas in the RN regime it is characterized by the mean
  field (MF) scaling. On SWN, both the scaling behaviors are found to
  coexist. Small compact avalanches below certain characteristic size
  $s_c$ are found to belong to the BTW universality class whereas
  large, sparse avalanches above $s_c$ are found to belong to the MF
  universality class. A scaling theory for the coexistence of two
  scaling forms on SWN is developed and numerically verified. Though
  finite size scaling (FSS) is not valid for DSM on RL as well as on
  SWN, it is found to be valid on RN for the same model. FSS on RN is
  appeared to be an outcome of super diffusive sand transport and
  uncorrelated toppling waves.
\end{abstract}

\pacs{89.75.-k,05.65.+b,64.60.aq}

\maketitle

\section{INTRODUCTION}
Complex networks describe a wide range of systems in nature and
society \cite{barbasi1}.  Frequently cited examples include coupled
biological and chemical systems \cite{jeong}, neural networks
\cite{anna}, internet \cite{falou}, world wide web \cite{barbasi},
social networks \cite{stanley}, networks of coauthors \cite{newmann},
citation network \cite{render}, wealth network \cite{garl}, etc.  A
small world network (SWN) introduced by Watt and Strogatz \cite{sw} is
a partially disordered structure interpolating between the regular
lattice (RL) and random network (RN). An SWN with a specified shortcut
density $\phi$, number of shortcuts per existing link, is generated
adding extra links (or shortcuts) between two randomly chosen sites of
the lattice keeping all the original bonds of the lattice intact
\cite{newman_book2}. In this process, $\phi=0$ corresponds to RL and
$\phi=1$ corresponds to a fully grown RN. A fully grown RN is
characterized by Poissonian degree distribution \cite{erenyi}. As
$\phi$ increases from $0$ there will an onset of small world behavior
around $\phi \approx 1/N$ where $N$ is the number of nodes present in
the network \cite{newman,newman_book2}. The small world behavior is
characterized by the fact that the shortest distance $\ell$ between
any two nodes is small as that of a RN and at the same time the
concept of neighborhood is preserved as that of a RL \cite{havlin}. If
$\phi$ is increased further, the small world behavior will evolve to
that of a RN around $\phi\approx 0.1$ \cite{newman_book2}. There exits
a characteristic length $\xi\sim \phi^{-1/d}$ where $d$ is the
dimensionality of the lattice, below which SWN belongs to ``large
world'' regime (RL) and beyond which it behaves as ``small world''
\cite{newman_rg,mendes00}.  Depending on the value of $\xi$, the
average shortest distance $\langle \ell \rangle$ scales with the
system size $L$ as
\begin{equation}
  \label{l_scale1} 
 \langle \ell \rangle = L\mathcal{F}(L/\xi) =
 L\mathcal{F}(\phi^{1/d}L)
\end{equation} 
where $\mathcal{F}(\phi^{1/d}L)$ is a universal scaling function
\cite{newman,barthelemy} and is given by
\begin{eqnarray}
\label{l_scale2}
 \mathcal{F}(x)&\propto& \left \{ \begin{array}{ll}
  \mbox{constant}, & x \ll 1
  \\(\log x)/x,  & x \gg 1 \\
       \end{array} \right.
\end{eqnarray}
It can be noted here that the scaling form was exactly determined for
one dimensional SWN by Newman {\em et al.} \cite{newman_ex} except for
$x=1$.

On the other hand, a commonly occurring phenomenon in nature and
society is self organized criticality (SOC) \cite{jen} which refers to
the intrinsic tendency of a wide class of slowly driven systems to
evolve spontaneously to a non equilibrium steady state characterized
by long range spatiotemporal correlation and power law scaling
behavior. SOC is observed in many physio-chemical process such as
earthquake \cite{ek}, forest fire \cite{ff}, biological evolution
\cite{be}, droplet formation \cite{df}, superconducting avalanches
\cite{sa}, etc. In order to study SOC, Bak, Tang and Weisenfeld (BTW)
\cite{btw} developed a simple lattice model called sandpile. The model
and its several different variants have been extensively studied on RL
and a large amount of analytical and numerical results were reported
in the literature \cite{dhar_rev}.

In nature, there exists many situations where self-organization occurs
in systems having complex structure such as network. For example
propagation of neural information inside the cervical cortex,
earthquake dynamics on the network of faults in the crust of the
earth, propagation of information through a network with
malfunctioning router and many others. Existence of such phenomena
triggered studies of SOC dynamics on complex networks in recent time
\cite{ebon,herrmann,graham,goh,kaski,marshili,holyst,optsc}. In this
paper, a generalized dissipative sandpile model (DSM) with variable
critical height is developed on a series of SWNs in order to examine
the effect of different length scales present in SWNs on the critical
behavior of sandpile dynamics as well as that of slowly driven
dynamical systems in general. It is found that the critical properties
of DSM on the RL ($\phi=0$) characterized by BTW type multiscaling
\cite{malcai} evolve to that of DSM on a RN ($\phi=1$) characterized
by mean field (MF) scaling \cite{chris}. For intermediate values of
$\phi$ $(0<\phi<0.1)$, coexistence of both the critical behavior is
observed corresponding to the presence of neighborhood as well as long
distance connectivity simultaneously on SWNs. In the following a
scaling theory for the coexistence scaling is proposed and verified by
extensive numerical simulation. It is also demonstrated that finite
size scaling (FSS) for BTW type models can only be valid if the system
on which the model is defined has no spatial structure {\em i.e.} on
RN and it will not be valid if the concept of neighborhood persists
{\em i.e.} on RL and SWN.

\section{DSM on SWN}
In this model, SWNs are generated by adding shortcuts between two
randomly chosen sites of a two dimensional ($2d$) square lattice of
size $L$. There are $L^2$ nodes and $2L^2$ bonds present on the $2d$
square lattice if periodic boundary condition is assumed. The number
of nodes is kept fixed to $L^2$ throughout the simulation. RL is
modified to SWN by adding shortcuts between two arbitrary sites of the
RL with a specified density $\phi$. The sites are chosen uniformly
from all over the lattice. The density of extra link per existing bond
of the original lattice is defined as 
\begin{equation}
\label{scd}
\phi=N_\phi/(2L^2) 
\end{equation}
where $N_\phi$ is number of shortcuts added to the lattice. Measure
has been taken to avoid more than one link between any two
nodes. There is no link which connects a node itself. Each $\phi$
value corresponds to a particular SWN.  $\phi=0$ corresponds to RL and
$\phi=1$ corresponds to a fully grown RN. It is verified for $\phi=1$
that the degree distribution of the network is given by a Poisson
distribution.

An SWN of a given $\phi$ is now driven by adding sand grains, one at a
time, to randomly chosen nodes. The height of the sand column of each
node is stored in an integer variable $h_i, i=1,2,\cdots,L^2$. For a
given $\phi$, the nodes of the SWN will have a particular degree
distribution. If the $i$th node has degree $k_i$, the critical height
or the threshold value for toppling of the $i$th node is taken to be
its degree $k_i$. If the height of the sand column at any node becomes
greater than or equal to the threshold value ($k_i$), it will be
marked as unstable. The corresponding sand column then topples and the
height of the sand column is reduced by its degree $k_i$. The node
then becomes under critical. The sand grains flow from the toppled
node to its adjacent nodes which are connected to the toppled node by
links.  Since there is no rigid boundary exists for a network, the
boundary sites of a RL where sand dissipation used to occur are
supposed to be distributed among randomly selected nodes of the
network. Dissipation of sand to those nodes is made with an
appropriate dissipation factor $\epsilon_\phi$ in an annealed
manner. It is realized by dissipating a sand grain with probability
$\epsilon_\phi$ in every attempt of sand transport from the critical
node. The adjacent nodes are then called sequentially one by one and
every time $\epsilon_\phi$ is compared with a random number $r$. If
$r\le\epsilon_\phi$, the sand grain is dissipated out from the system
and the height of the sand column at the corresponding adjacent node
remains the same otherwise it is increased by one unit. The toppling
rule then can be represented as
\begin{equation}
\label{trule2}
\begin{array}{ll}
h_i \rightarrow  h_i-k_i &\\
 h_j \rightarrow h_j & \hspace{0.5cm} \mbox{if} \hspace{0.2cm} r \le
 \epsilon_\phi, \\ h_j \rightarrow h_j+1 & \hspace{0.5cm}
 \mbox{otherwise}
\end{array}
\end{equation}
where $j = 1,2,3\cdots k_i$. If the toppling of a node causes some of
the adjacent nodes unstable, subsequent toppling follow on these
unstable nodes. The process continues until there is no unstable node
present in the system. These toppling activities lead to an
avalanche. During an avalanche no sand grain is added to the system.

The critical properties of DSM are studied on SWNs defined on the
square lattice of different sizes varying $\phi$ from $0$ to $1$ for
each lattice size. It is now essential to determine the dissipation
factor $\epsilon_\phi$ for an SWN of given $\phi$ and system size $L$.

\section{Determination of $\epsilon_\phi$}
\begin{figure}[t]
\centerline{\hfill
  \psfig{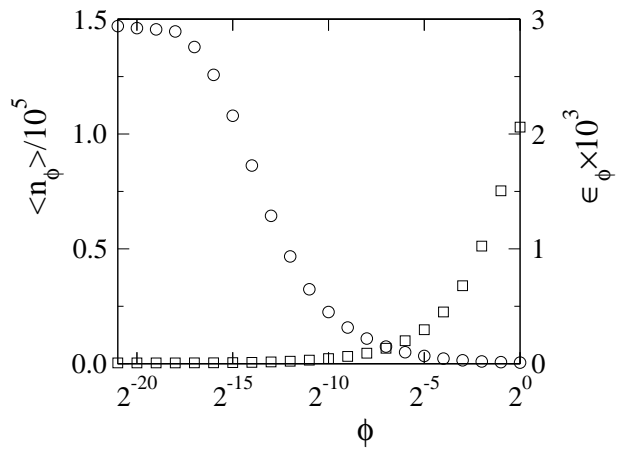} \hfill}
\caption{Plot of $\langle n_\phi \rangle$ ($\circ$) and
  $\epsilon_\phi$ ($\Box$) against $\phi$ in log-normal scale for
  $L=1024$. }
\label{n_phi}
\end{figure} 

Malcai {\em et al.} \cite{malcai} defined the dissipation factor for a
DSM on RL by the inverse of time steps required for a random walker
to reach the lattice boundary starting from an arbitrary site. Such a
definition for the dissipation factor on RL has been extend to SWN
here. The dissipation factor $\epsilon_\phi$ on an SWN corresponding
to a given $\phi$ is then given by
\begin{equation}
  \label{ng1} 
  \epsilon_\phi = \frac{1}{\langle n_\phi \rangle}
\end{equation}  
where $\langle n_\phi \rangle$ is the average number of steps required
for a random walker to reach the lattice boundary starting from an
arbitrary node of the SWN. The average number of steps $\langle n_\phi
\rangle$ required for such walks is calculated by performing $2 \times
10^6 $ random walks in $16$ different random configuration of every
SWN. In performing such walks no periodic boundary condition is
applied. In Fig.\ref{n_phi}, $\langle n_\phi \rangle$ and
$\epsilon_\phi$ are plotted against $\phi$ in semi-logarithmic scale
for $L=1024$. It can be seen that $\langle n_\phi \rangle$ decreases
rapidly with increasing $\phi$. This is because as $\phi$ increases
the number of shortcuts also increases in the system and consequently
the walker needs lesser number of step to reach the boundary starting
from an arbitrary node. Consequently, $\epsilon_\phi$ increases
rapidly as $\phi\rightarrow 1$. For the two extreme values of $\phi$,
the dissipation factors are obtained as
$\epsilon_{\phi=0}=6.7\times10^{-6}$ and $\epsilon_{\phi=1}=0.002$.

Using the estimated $\epsilon_\phi$, sandpile dynamics now can be
studied on SWNs at different $\phi$ on a given $L$.

\section{Steady State of DSM on SWN}
\begin{figure}[t]
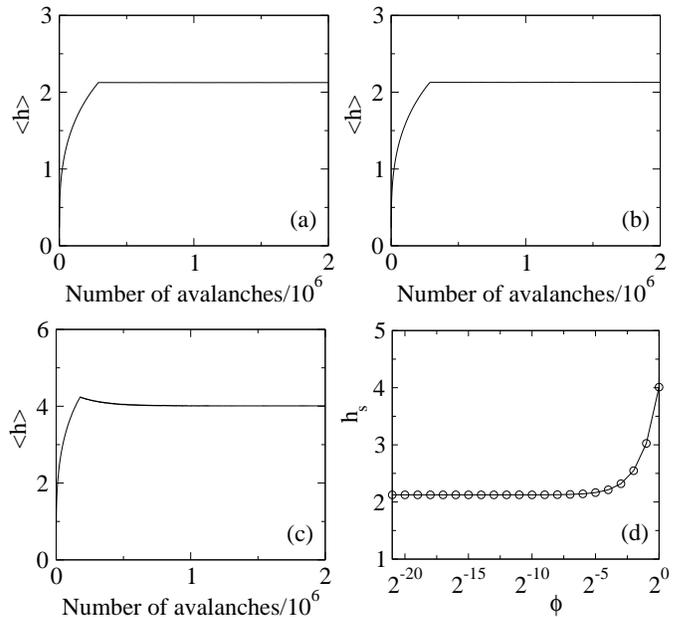

\centerline{\hfill
  \psfig{file=santra_fig_2a.eps,width=0.24\textwidth} \hfill
  \psfig{file=santra_fig_2b.eps,width=0.24\textwidth} \hfill
  }
\vspace{0.2cm}
\centerline{\hfill
  \psfig{file=santra_fig_2c.eps,width=0.24\textwidth} \hfill\hfill
    \psfig{file=santra_fig_2d.eps,width=0.245\textwidth} \hfill
   \hfill}
\caption{\label{avht} Plot of average height $\langle h \rangle$
  against number of avalanches of DSM on SWNs generated on a square
  lattice of size $L=1024$ for $\phi=0$ (a), $\phi=2^{-8}$ (b), and
  $\phi=1$(c). In (d), plot of average saturated height $h_s$ against
  $\phi$ in log-normal scale. }
\end{figure}

The steady state of DSM on an SWN corresponds to equal current of
incoming flux of sand grains into the system to that of outgoing flux
of the sand grains from the system. Thus, at the steady state
condition the average height $\langle h \rangle$ of the sand columns
should remain constant. For $L^2$ nodes, the average height is defined
as
\begin{equation}
\label{st} 
\langle h \rangle = \frac{1}{L^2}\sum_{i=1}^{L^2}h_i.
\end{equation}  
In Fig.\ref{avht}, $\langle h \rangle$ is plotted against the number
of avalanches for SWNs defined on a square lattice of size $L=1024$
for $\phi=0$ (a), $ 2^{-8}$ (b) and $1$ (c). It can be seen that the
steady state for DSM is achieved after initial $10^6$ avalanches in
all the SWNs considered. The saturated average height $h_s$ is plotted
against $\phi$ in Fig.\ref{avht}(d). The value of $h_s$ on the regular
lattice, $\phi=0$, is approximately $2.125$ as it was conjectured in
the context of absorbing state phase transition of fixed energy
sandpile model on the square lattice \cite{vespignani}. As $\phi$
increases, the value of $h_s$ remains almost independent of $\phi$
upto $\phi\approx 2^{-3}$ and beyond this value of $\phi$, $h_s$
increases rapidly with $\phi$. Since the avergae critical height of
the sandpile model on an SWN is defined by the average degree $\langle
k \rangle$ of the network, the variation of $h_s$ with $\phi$ must be
due to the change of $\langle k \rangle$ with $\phi$. A simple
relationship between $\langle k \rangle$ and $\phi$ can be obtained as
\begin{equation}
\label{kphi} 
\langle k \rangle = 4+2\frac{2L^2\phi}{L^2}=4(1+\phi)
\end{equation}
where $2L^2\phi$ is the number of shortcuts added and the factor $2$
corresponds to increase of degree by one of two nodes for addition of
each shortcut. Therefore, for $\phi=0$, $\langle k \rangle=4$, for
$\phi=1$, $\langle k \rangle=8$ and for $\phi=2^{-3}$, $\langle k
\rangle=4.5$. Thus upto $\phi=2^{-3}$, the increase in $\langle k
\rangle$ is small because for $\phi<0.1$ the network corresponds to
the small world regime and the concept of neighborhood is
preserved. Since $\langle k \rangle$ is small in this region, the
change in $h_s$ is expected to be small. For $\phi>0.1$, $\langle k
\rangle$ increases rapidly and hence the value of $h_s$. It is also
observed that as $\phi$ increases, the steady state appear after an
initial hump in $\langle h \rangle$. For large $\phi$, the dissipation
in the system will be mostly through the nodes with higher degrees. It
takes some time for those nodes to accumulate appropriate number of
sand grains to become critical. During the initial piling up of the
sand columns in the higher degree nodes, the average height of the
sand columns may increase beyond the saturation value $h_s$
corresponding to the steady state.

\section{Probability distributions and conditional expectation values of avalanche properties}

The critical behavior of different avalanche properties like toppling
size $s$, area $a$ and lifetime $t$ of an avalanche are measured to
characterize the DSM on SWNs. The toppling size $s$ is defined as the
total number of toppling which occurred in an avalanche, the avalanche
area $a$ is equal to the number of distinct sites or nodes toppled in
an avalanche and the lifetime $t$ of an avalanche is the number of
parallel updates to make all the nodes (sites)
under-critical. Sandpile dynamics is mostly characterize the
probability distribution of these avalanche properties and the
conditional expectation values $\langle x_\phi(y)\rangle$ of a
property $x$ keeping another property $y$ fixed at a certain value
\cite{chris_conexp}. At the steady state, the probability distribution
functions $P(x,\phi)$ on SWN generated on a large lattice of fixed
size with a given $\phi$ is expected to obey power law scaling as
\begin{equation}
\label{pd}
P(x,\phi) \sim x^{-\tau_x(\phi)}
\end{equation}
where $x\in \{s,a,t\}$ and $\tau_x(\phi)$ is the critical exponent
corresponding to the given value of $\phi$. The conditional
expectation $\langle x_\phi(y)\rangle$ is defined as
\begin{equation}
\label{cpd}
\langle x_\phi(y)\rangle = \int_{0}^{x_{max}} xP(x|y)dx
\end{equation}
where $P(x|y)$ is the conditional probability of the property $x$ for
a fixed value of $y$. The quantity $\langle x_\phi(y)\rangle$ is expected
to scale with the other property $y$ as
\begin{equation}
\label{cev}
\langle x_\phi(y)\rangle \sim y^{\gamma_{xy}(\phi)}
\end{equation}
where $x\in \{s,a,t\}$ and $\gamma_{xy}(\phi)$ is another $\phi$
dependent critical exponent. The exponent $\gamma_{xy}(\phi)$ can also
be obtained in terms of the distribution exponents $\tau_x(\phi)$ and
$\tau_y(\phi)$ as given in \cite{rsm}, 
\begin{equation}
\label{scgt}
\gamma_{xy}(\phi) = \frac{\tau_y(\phi)-1}{\tau_x(\phi)-1}.
\end{equation}

Before analyzing the probability distributions and the conditional
probabilities, one should notice that there exists a length scale
$\xi$ for a given SWN below which the SWN behaves as RL and above
which it behaves as network. It is then expected that there should
exist a characteristic value $x_c$ of every avalanche property
corresponding to the length scale $\xi$ of SWN. For a given $\phi$,
below and above $x_c$ the probability distributions and the
conditional probabilities are then expected to behave differently. In
two dimensions, $\xi$ scales with $\phi$ as $\xi\sim\phi^{-1/2}$
\cite{newman}. Therefore, the characteristic area $a_c$ of the
avalanches occurring on RL must be proportional to $\xi^2$. Hence, the
sacling of $a_c$ with $\phi$ should be given by
\begin{equation}
\label{ac}
a_c \sim \phi^{-\alpha_a}
\end{equation}
with $\alpha_a=1$. Knowing the scaling of $a_c$ with $\phi$, one can
find the scaling of $s_c$ and $t_c$ with $\phi$ as well. From the
conditional expectation of avalanche size for fixed avalanche area one
expects $s_c \sim a_c^{\gamma_{sa}}$ on RL. Hence,
\begin{equation}
\label{sc}
s_c \sim \phi^{-\gamma_{sa}}.
\end{equation}
Then, $\alpha_s=\gamma_{sa}$. Similarly, $t_c \sim
s_c^{\gamma_{ts}}$ or  $t_c \sim
a_c^{\gamma_{sa}/\gamma_{st}}$. Therefore, one has
\begin{equation}
\label{tc}
t_c  \sim \phi^{-\gamma_{sa}/\gamma_{st}}.
\end{equation}
Hence, $\alpha_t=\gamma_{sa}/\gamma_{st}$. Since $\gamma_{sa}=1.06$
and $\gamma_{st}=1.63$ on RL \cite{rsm}, the values of $\alpha_s$ and
$\alpha_t$ are expected to be $1.06$ and $0.65$ respectively.

\begin{figure}[t]
\centerline{\hfill
\psfig{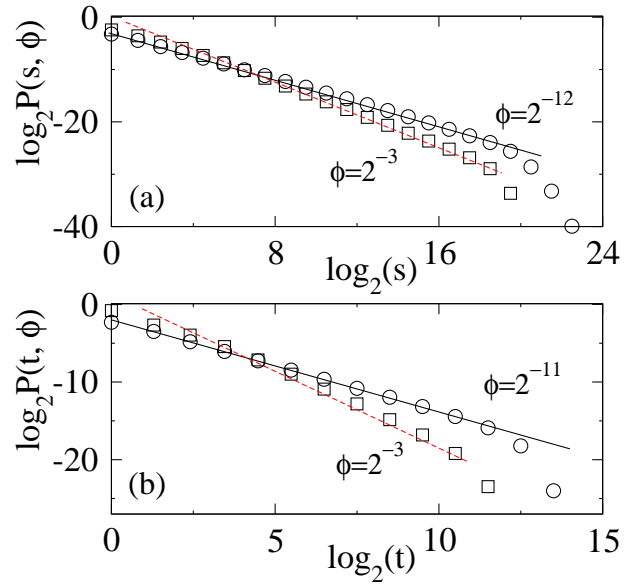}\hfill}
\caption{(Color online) (a) Distribution of avalanche size $P(s,\phi)$
  is plotted in double logarithmic scale for two different values of
  $\phi$: $\phi=2^{-12}$ ($\Circle$) and $\phi=2^{-3}$
  ($\Box$). Straight lines are fitted through data points. The slope
  of the solid black line is $-1.11 \pm 0.01$ and that of the dashed
  red line is $-1.50 \pm 0.01$. (b) Plot of $P(t,\phi)$ against
  $t$. The slope of the solid black line is $-1.18 \pm 0.01$ and the
  dashed red line has slope $-1.98 \pm 0.02$. }
\label{pds}
\end{figure}
\begin{figure}[t]
\centerline{\hfill
\psfig{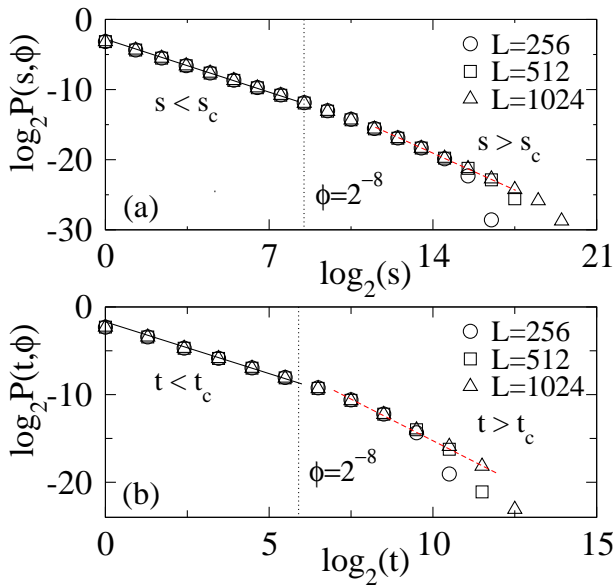}\hfill}
\caption{(Color online) (a) Distribution of avalanche size $P(s,\phi)$
  is plotted in double logarithmic scale for $\phi=2^{-8}$ for three
  different system size $L=256$($\Circle$), $512$($\Box$),
  $1024$($\triangle$). Vertical dotted line at $s=s_c$ seperates the
  two regimes. The slope of the solid black line for $s<s_c$ is $-1.11
  \pm 0.01$ and that of the dashed red line for $s>s_c$ is $-1.50 \pm
  0.01$. (b) Plot of $P(t,\phi)$ versus $t$. Vertical dotted line at
  $t=t_c$ seperates the two regimes. The black solid line for $t<t_c$
  has the slope $-1.18\pm0.01$ and red dashed line for $t>t_c$ has
  slope $-1.98\pm0.02$.}
\label{pds2}
\end{figure}
\begin{figure}[t]
\centerline{\hfill
\psfig{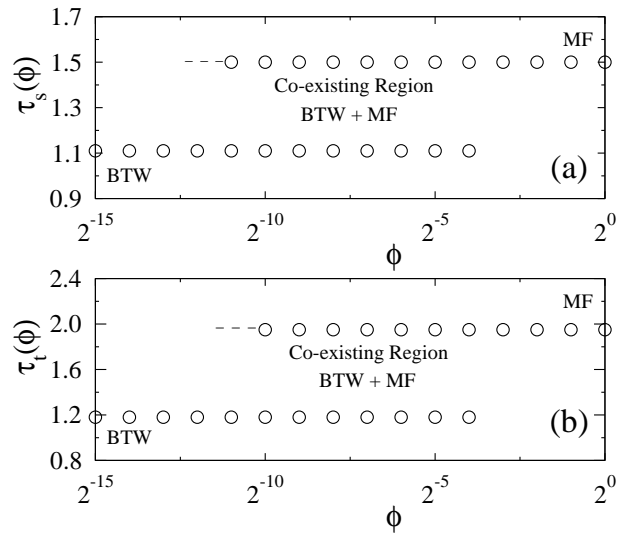}\hfill}
\caption{The values of the exponent $\tau_s(\phi)$ against $\phi$ are
  given in (a) and that of $\tau_t(\phi)$ against $\phi$ are given in
  (b). The extrapolated dashed lines indicate possible exponent values
  for larger system sizes.}
\label{pds3}
\end{figure}

In order to estimate the probability distributions of the avalanche
properties and their conditional expection values, the following
statistical averages are made. Sixteen SWNs configurations are
considered for a given $\phi$. On each SWN, after attaining the steady
state $10^6$ avalanches are neglected and next $2\times 10^6$
avalanches are collected. Therefore, a total of $32\times 10^6$
avalanches are taken for data averaging. The value of $\phi$ is varied
from $0$ to $1$ increasing $N_\phi$ in multiple of $2$.  The values of
$\tau_x(\phi)$ are estimated determining the probability distributions
of the respective avalanche properties $x\in \{s,a,t\}$.

Probability distributions of avalanche size $P(s,\phi)$ and that of
lifetime $P(t,\phi)$ are estimated on SWNs generated for different
values of $\phi$ for a given lattice size. For $\phi=2^{-12}$ (close
to $\phi=0$) and $\phi=2^{-3}$ (close to $\phi=1$), $P(s,\phi)$ and
$P(t,\phi)$ obtained on a lattice of size $L=1024$ are plotted in
Fig.\ref{pds}(a) and Fig.\ref{pds}(b) respectively. It can be seen
that for both $\phi=2^{-12}$ as well as for $\phi=2^{-3}$, $P(s,\phi)$
and $P(t,\phi)$ have power law behavior almost over the whole extent
of $s$ and $t$ but with different critical exponents. The scaling
behavior at $\phi=2^{-12}$ is found to be characterized by the
avalanche size exponent $\tau_s=1.11 \pm 0.01$, and avalanche time
exponent $\tau_t=1.18 \pm 0.01$, which are measured by the best fitted
straight line (black) through the data points. The value of $\tau_s$
for $\phi=2^{-12}$ is same as that of previously reported for DSM on
RL ($\phi=0$) \cite{shilo,malcai}. Note that the value of $\tau_s$ of
the BTW model (Dhar abelian sandpile model \cite{dhar_rev}) was also
reported to be $\approx 1.11$ though in the $L\rightarrow \infty$
limit it is expected to be $\approx 1.29$
\cite{solomon,lubeck,lubeck_ex}. Therefore the BTW type sandpile
dynamics on RL remains unperturbed when performed on a lattice with
additional $N_\phi=512$ shortcuts corresponding to $\phi=2^{-12}$ on a
lattice of size $L=1024$. On the other hand, the power law scaling at
$\phi=2^{-3}$ is found to be characterized by a critical exponents
$\tau_s= 1.5 \pm 0.01$ and $\tau_t= 1.98 \pm 0.02$. The value of
$\tau_s$ for DSM obtained by mean field (MF) theory for lattices
without spatial structure \cite{chris} as well as by branching process
for RN \cite{chris,ebon} was known to be $3/2$ and the exact value of
$\tau_t$ on RN obtained by branching process is $2$
\cite{ebon}. Hence, the measured value of $\tau_t=1.98$ for RN is
close to the exact result. Therefore by the addition of
$N_\phi=2^{18}$ shortcuts corresponding to $\phi=2^{-3}$ on a lattice
of size $L=1024$, the RL evolves to a RN and DSM scaling on it can be
described by MF scaling though the critical height is not taken as the
mean degree of nodes as it was taken in Ref.\cite{chris}.

For $\phi=2^{-8}$ (an intermidiate value of $\phi$), $P(s,\phi)$ and
$P(t,\phi)$ are plotted in Fig.\ref{pds2}(a) and Fig.\ref{pds2}(b)
respectively for different values of system size $L$. It is
interesting to note that for $\phi=2^{-8}$, the distributions of
$P(s,\phi)$ and $P(t,\phi)$ follow two different power law scaling at
different regimes of $s$ and $t$ respectively, separated by a
characteristic value $s_c$ and $t_c$ as shown by dotted lines in
Fig.\ref{pds2}(a) and Fig.\ref{pds2}(b). The values of $s_c$ and $t_c$
are obtained from Eq.\ref{sc} and Eq.\ref{tc} respectively. For
$s<s_c$, the scaling behavior of $P(s,\phi)$ is characterized by
$\tau_s = 1.11 \pm 0.01$ whereas for $s>s_c$ it is characterized by
$\tau_s = 1.5 \pm 0.01$. Therefore for SWNs corresponding to
intermediate values of $\phi$, both the scaling forms, BTW and MF, of
DSM coexist. It should also be noticed that the characteristic size
$s_c$ or characteristic time $t_c$ does not change with system size
$L$, as the characteristic length scale $\xi$ does not depend on $L$
\cite{newman}. The values of the critical exponent $\tau_s(\phi)$ and
$\tau_t(\phi)$ obtained for different values of $\phi$ are given in
Fig.\ref{pds3}(a) and Fig.\ref{pds3}(b). It is important to note that
coexistence of both the scaling forms persists over a wide range of
$\phi$ given by $2^{-12} < \phi < 2^{-3}$ for $s$ and $2^{-11} < \phi
< 2^{-3}$ for $t$. The upper limit corresponds to crossover of SWN to
RN at $\phi\approx 0.1$ \cite{newman_book2}. Though the crossover from
RL to SWN occurs at $\phi\approx 1/L^2$ \cite{newman}, for the finite
system of size $L=1024$ the sandpile dynamics is able to recognize
such a crossover only at $\phi=2^{-12}$ (or $\sim 10^{-4}$). If the
system size increases, such crossover is expected to appear in the
sandpile dynamics for smaller values of $\phi$ and both exponents
would be possible to measure in this regime as shown by a dashed line
in Fig.\ref{pds}(c). The avalanche area $a$ also displays a similar
co-existence of scaling behavior over the same range of $\phi$. For
RL, $\tau_a$ is found $1.12$ as per the reported value for the BTW
model for finite systems \cite{lubeck_ex}. However for RN, it is found
that the value of $\tau_a=1.5$ as that of $\tau_s$ on RN. In the SWN
regime, both the scaling forms are found to coexist. 

\begin{figure}[t]
\centerline{\hfill
 \psfig{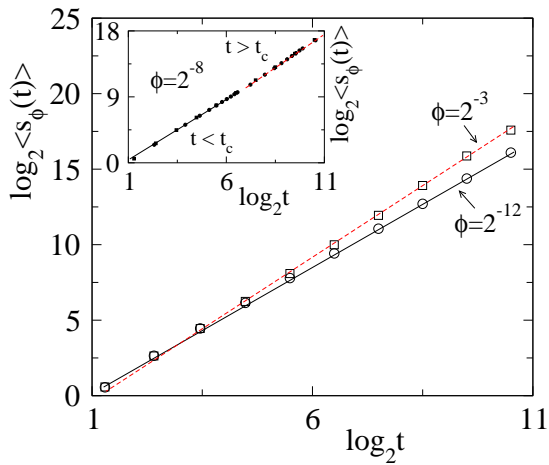}\hfill
}
\caption{(Color online) Plot of $\langle s_\phi(t)\rangle$ against $t$
  for $\phi=2^{-12}$ ($\circ$) and $\phi=2^{-3}$ ($\Box$). Solid lines
  are the best fitted straight lines having slope $1.62\pm0.02$ (black
  solid line) for $\phi=2^{-12}$ and $1.98\pm0.03$ (red dashed line)
  for $\phi=2^{-3}$. In the inset $\langle s(t)\rangle$ against $t$
  for $\phi=2^{-8}$ is plotted. }
\label{gamma_st}
\end{figure}

The coexistance of scaling is also verified for the condiotional
expectation value $\langle s_\phi(t)\rangle$. Its variation against
$t$ for $\phi=2^{-12}$ and $2^{-3}$ are shown in
Fig.\ref{gamma_st}. The critical exponent $\gamma_{st}$ are obtained
as $1.62\pm 0.02$ and $1.98\pm 0.03$ for $\phi=2^{-12}$ and
$\phi=2^{-3}$ respectively. Since on RL $\tau_s=1.11$ and
$\tau_t=1.18$, the expected value of $\gamma_{st}$ from the scaling
relation Eq.\ref{scgt} is $1.63$ on RL. Similarly for RN, $\tau_s=1.5$
and $\tau_t=2$, the expected values of $\gamma_{st}=2$ on RN. The
values of the critical exponents $\gamma_{st}$ are within the error
bars of the expected values. In the inset of Fig.\ref{gamma_st},
$\langle s(t)\rangle$ is plotted against $t$ for $\phi=2^{-8}$. Two
different scaling of $\langle s_\phi(t)\rangle$ with $t$ are shown by
back solid line and red dashed line respectively for $t<t_c$ and
$t>t_c$. The coexistence scaling of $\langle s_\phi(t)\rangle$ is also
observed for the same range of $\phi$ as it was observed for the
avalanche size distribution.

It could be recalled that in an SWN there exists the concept of
neighborhood corresponding to RL at the same time the shortest
distance $\ell$ between two nodes is vanishingly small corresponding
to RN. Because of the coexistence of both the characteristics of RL as
well as that of RN in an SWN, the sandpile avalanches are segregated
according to their sizes into two scaling forms. It should be
emphasized here that such coexistence of two scaling behaviors is also
observed on SWNs generated by removing the bonds emanating from a site
of a square lattice with probability $\phi$ and rewiring it to a
randomly selected lattice site. However, in contrary to the present
observation, Arcangelis and Hermann \cite{herrmann} obtained a
continuous crossover from BTW universality class to MF universality
class in the study of a BTW type sandpile dynamics on SWNs constructed
by rewiring a fraction of bond of a square lattice keeping the
critical height same for all the nodes and having dissipation only at
the lattice boundary. No coexisting region of both the scaling forms
was observed in their study. On the other hand, in the study of one
dimensional sandpile model on SWNs a transition from non-critical to
critical regime was demonstrated by Lahtinen {\em et al.}
\cite{kaski}.

\begin{figure}[b]
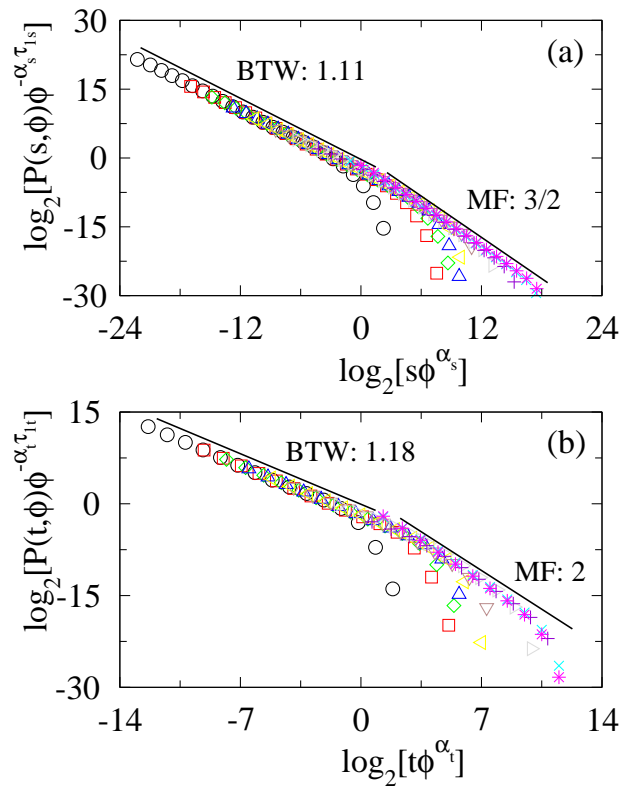

\centerline{\hfill
  \psfig{file=santra_fig_7a.eps,width=0.45\textwidth}\hfill}
\centerline{\hfill
  \psfig{file=santra_fig_7b.eps,width=0.45\textwidth}\hfill}
\caption{(Color online) (a) Plot of scaled distribution
  $P(s,\phi)\phi^{-\alpha_s\tau_{1s}}$ against a scaled variable
  $z_s=s\phi^{\alpha_s}$. (b) Plot of
  $P(t,\phi)\phi^{-\alpha_t\tau_{1t}}$ against
  $z_t=t\phi^{\alpha_t}$. Different symbols corresponding to different
  $\phi$ values are taken as: $\phi=2^{-21}(\Circle)$,
  $2^{-16}(\Box)$, $2^{-14}(\Diamond)$,
  $2^{-12}(\triangle)$,$2^{-10}(\triangleleft)$,
  $2^{-8}(\triangledown)$, $2^{-6}(\triangleright)$, $2^{-4}(+),
  2^{-1}(\times)$, $2^0(\hexstar)$.  Reasonable data collapse for both
  $s$ and $t$ are observed. Solid lines with respective slopes are
  guide to eye. }
\label{pdcoll}
\end{figure}

\section{Scaling of coexisting probability distributions}
As per the scaling form of the characteristic toppling
area $a_c$, size $s_c$, lifetime $t_c$ (obtained in Eqs.\ref{ac},\ref{sc}
and \ref{tc} respectively), a general scaling of the charateristic
property $x_c$ with $\phi$ is assumed as
\begin{equation}
\label{xc}
x_c(\phi) \approx \phi^{-\alpha_x}
\end{equation}
where $\alpha_x$s correspond to different characteristic exponents.
The values of $x_c(\phi)$ on RL ($\phi=0$), must correspond to the cut
off value of the distribution $P(x,0)$ for a given system size $L$. As
the network grows, the distribution $P(x,\phi)$ will develop a part
corresponding to MF scaling. Consequently the part representing the
BTW type scaling will shrink. Hence, the value of $x_c$ should
decrease with increasing $\phi$. Eventually, the value of $x_c$ will
be the one on RN when $\phi=1$. The existence of such a characteristic
value of toppling size as a function of was noticed in the sandpile
dynamics on one dimensional SWN \cite{kaski}.  It is now possible to
obtain a single probability distribution function for both the scaling
forms for the whole range of $\phi$.

A new scaling form for the distribution functions with respect to the
characteristic value $x_c(\phi)$ is now proposed as
\begin{equation}
\label{scaling}
P(x,x_c(\phi)) = \left \{ \begin{array}{ll} x^{-\tau_{1x}} {\sf
    f}\left(\frac{x}{x_c(\phi)}\right) & \mbox{for $x\leqslant x_c$}
  \\ x^{-\tau_{2x}} {\sf g}\left(\frac{x}{x_c(\phi)}\right) & \mbox{for
    $x\geqslant x_c$ }
       \end{array} \right. 
\end{equation}
where ${\sf f}$ and ${\sf g}$ are two different scaling functions in
two different regions and $\tau_{1x}$ and $\tau_{2x}$ are the
corresponding critical exponents in the respective regions. Since at
$x=x_c(\phi)$ the values of $P(x,x_c(\phi))$ are same for both the
regions, then one should have ${\sf f}(1) =
\phi^{-(\tau_{1x}-\tau_{2x})\alpha_x}{\sf g}(1)$. The probability
distribution then can be obtained in terms of a single scaling
function ${\sf f}$ or ${\sf g}$ as
\begin{eqnarray}
\label{scaling2}
P(x,\phi) &=& \left \{ \begin{array}{ll} x^{-\tau_{1x}} {\sf f}(x\phi^{\alpha_x})
  & \mbox{for $x\leqslant x_c$}
  \\ x^{-\tau_{2x}}\phi^{-\Delta_x\alpha_x} {\sf f}(x\phi^{\alpha_x})
  &\mbox{for $x\geqslant x_c$ }\\
       \end{array} \right. 
\end{eqnarray}
where $\Delta_x =(\tau_{2x}-\tau_{1x})$. The $\phi$ independent
scaling form can be obtained by rescaling the probability
distribution as
\begin{eqnarray}
\label{scaling3}
P(x,\phi)\phi^{-\alpha_x\tau_{1x}} &=& \left \{ \begin{array}{ll}
  z_x^{-\tau_{1x}} {\sf f}(z_x) & \mbox{for $z_x \leqslant 1$}
  \\ z_x^{-\tau_{2x}} {\sf f}(z_x) & \mbox{for $z_x \geqslant 1$ }\\
       \end{array} \right. 
\end{eqnarray}
where $z_x=x\phi^{\alpha_x}$ is a scaled variable. Such scaling
behavior was also observed in context of anomalous roughening of
fractured surface \cite{lopez}.

The above scaling forms are now verified. The rescaled probabilities
$P(x,\phi)\phi^{-\alpha_x\tau_{1x}}$ are plotted against the scaled
variable $z_x=x\phi^{\alpha_x}$ for $s$ and $t$ in Fig.\ref{pdcoll}(a)
and (b) respectively. It can be seen that a good data collapse is
obtained for both $s$ and $t$ using $\alpha_s=1.06$ and
$\alpha_t=0.65$. The critical exponents $\tau_x$ corresponding to two
different regions are also verified. The straight lines with required
slopes in the respective regions are guide to eye. It confirms the
proposed scaling form of the probability distribution functions on
SWNs. Such coexistence scaling in the SWN regime is also verified for
a stochastic sandpile model \cite{ssm}.

\begin{figure}[t]
\centerline{\hfill
\psfig{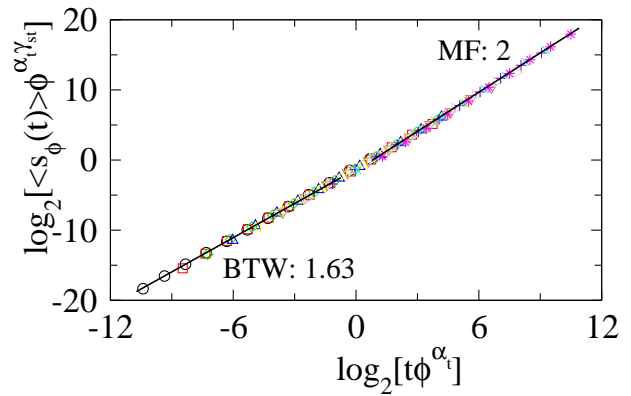}\hfill}
\caption{(Color online) Plot of $\langle s_\phi(t) \rangle
  \phi^{\alpha_t\gamma_{st}}$ against a scaled variable
  $t\phi^{\alpha_t}$. The same symbol set of Fig.\ref{pdcoll} is
  used. A good data collapse is observed. Two solid lines having
  slopes $1.63$ and $2$ indicating two different scaling forms are
  guide to eye.}
\label{st_coll}
\end{figure}

Since the probability distributions are now represented by a single
scaling form, the average avalanche properties can also be scaled in a
similar fashion. For example, the average cluster size is now expected
to scale as
\begin{equation}
\label{stphi}
\langle s_\phi(t)\rangle = t^{\gamma_{st}}f_{st}(t\phi^{\alpha_t})
\end{equation}  
where $f_{st}$ is a new scaling function and the value of
$\gamma_{st}$ correspond to that on RL. The form of the scaling
function is verified by plotting $\langle s_\phi(t)\rangle
\phi^{\gamma_{st}\alpha_t}$ against the scaled variable
$t\phi^{\alpha_t}$ in Fig.\ref{st_coll} taking $\gamma_{st}=1.63$. It
can be seen that there is a good data collapse and the scaling
function represents two different scaling behaviors with two different
exponents as $1.63$ and $2$, indicated by straight lines with
respective slopes. Such scaling behavior can also be obtained between
avalanche size $s$ and area $a$. Since $\alpha_a=1$ and
$\gamma_{sa}=1.06$, the change in slope in the scaling function is
difficult to observe in the numerical data collected here.

It is now important to understand the origin of coexistence of both
the critical behaviors of the avalanche properties on an SWN. Since
for an avalanche property $x$ there is BTW type scaling for $x<x_c$
and MF type scaling for $x>x_c$, it is intriguing to look into
the avalanche cluster morphology for the avalanches following two
different scaling behaviors.

\section{Avalanche cluster morphology}

\begin{figure}[t]
\centerline{\hfill
\psfig{file=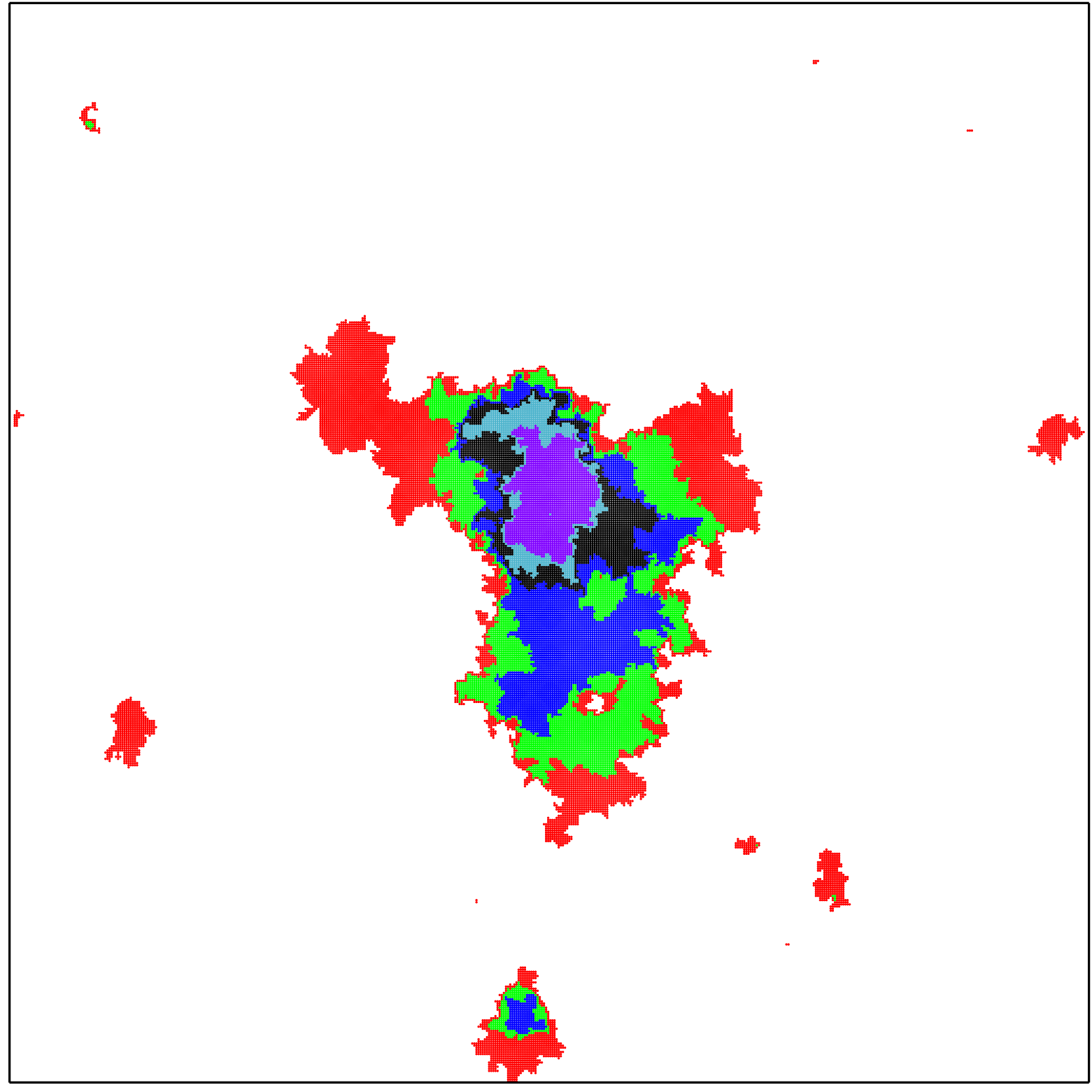,width=0.22\textwidth}\hfill
\psfig{file=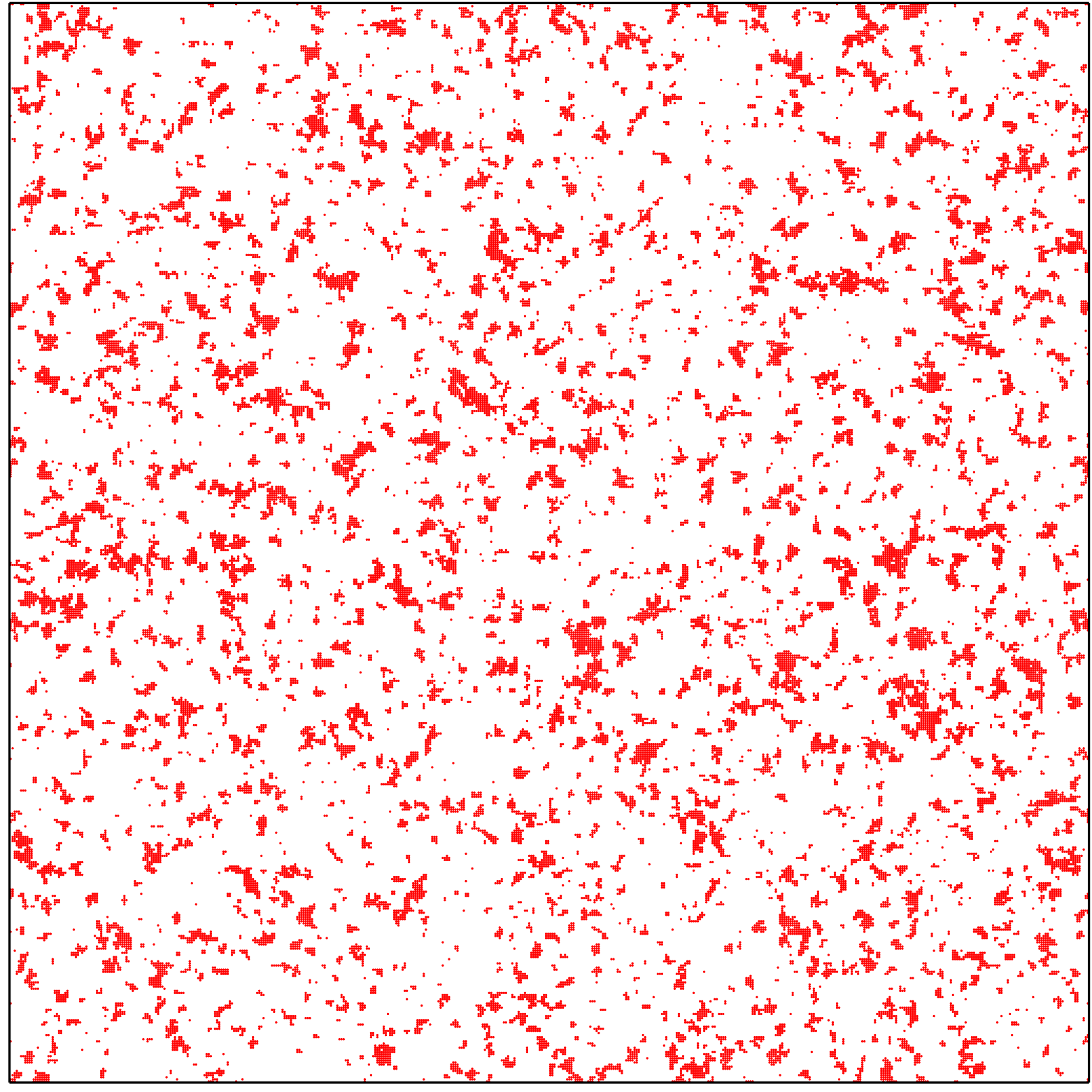,width=0.22\textwidth}\hfill}
\centerline{\hfill (a) $\phi=2^{-12}$ \hfill\hfill (b) $\phi=2^{-3}$ \hfill}
\medskip
\centerline{\hfill
  \psfig{file=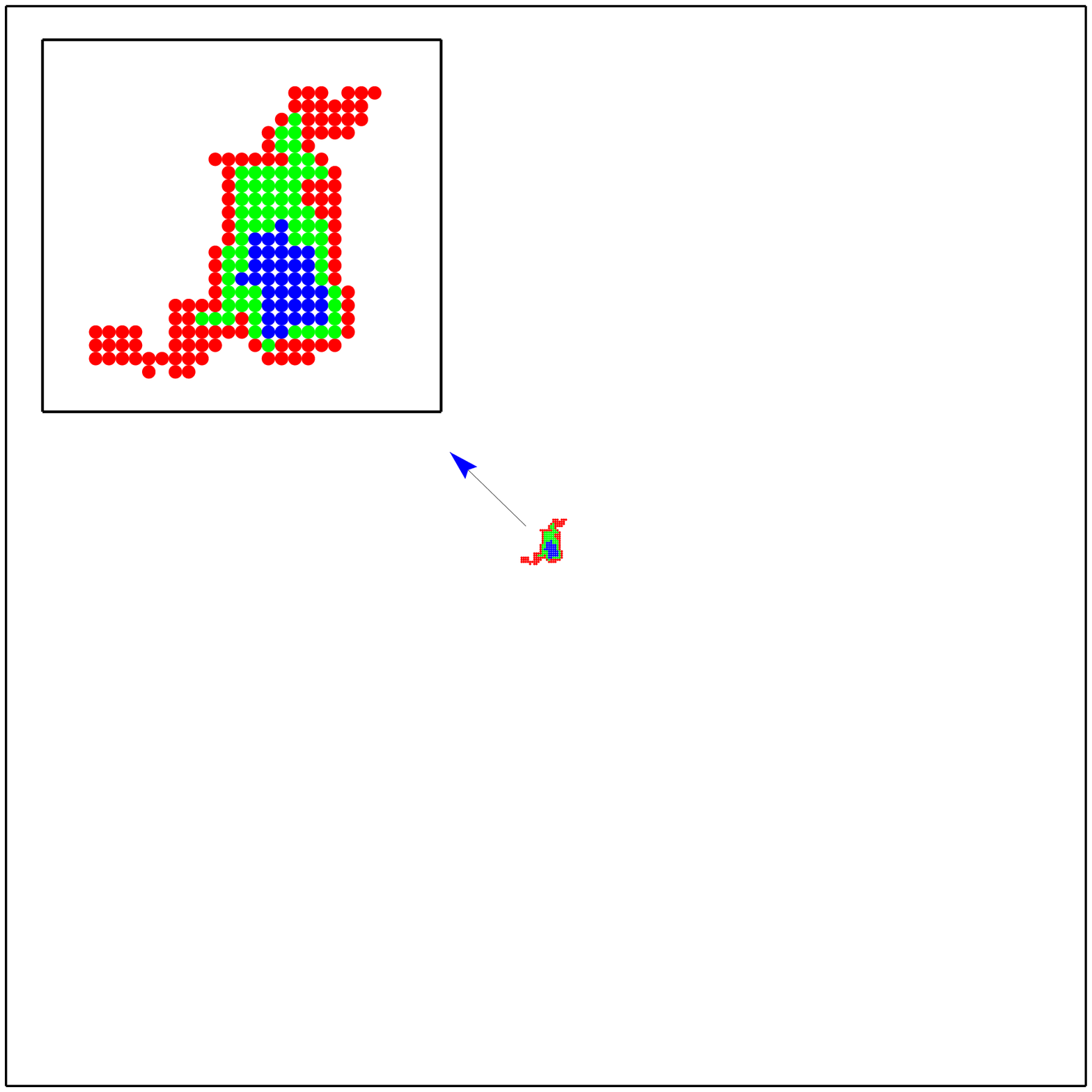,width=0.22\textwidth}\hfill
  \psfig{file=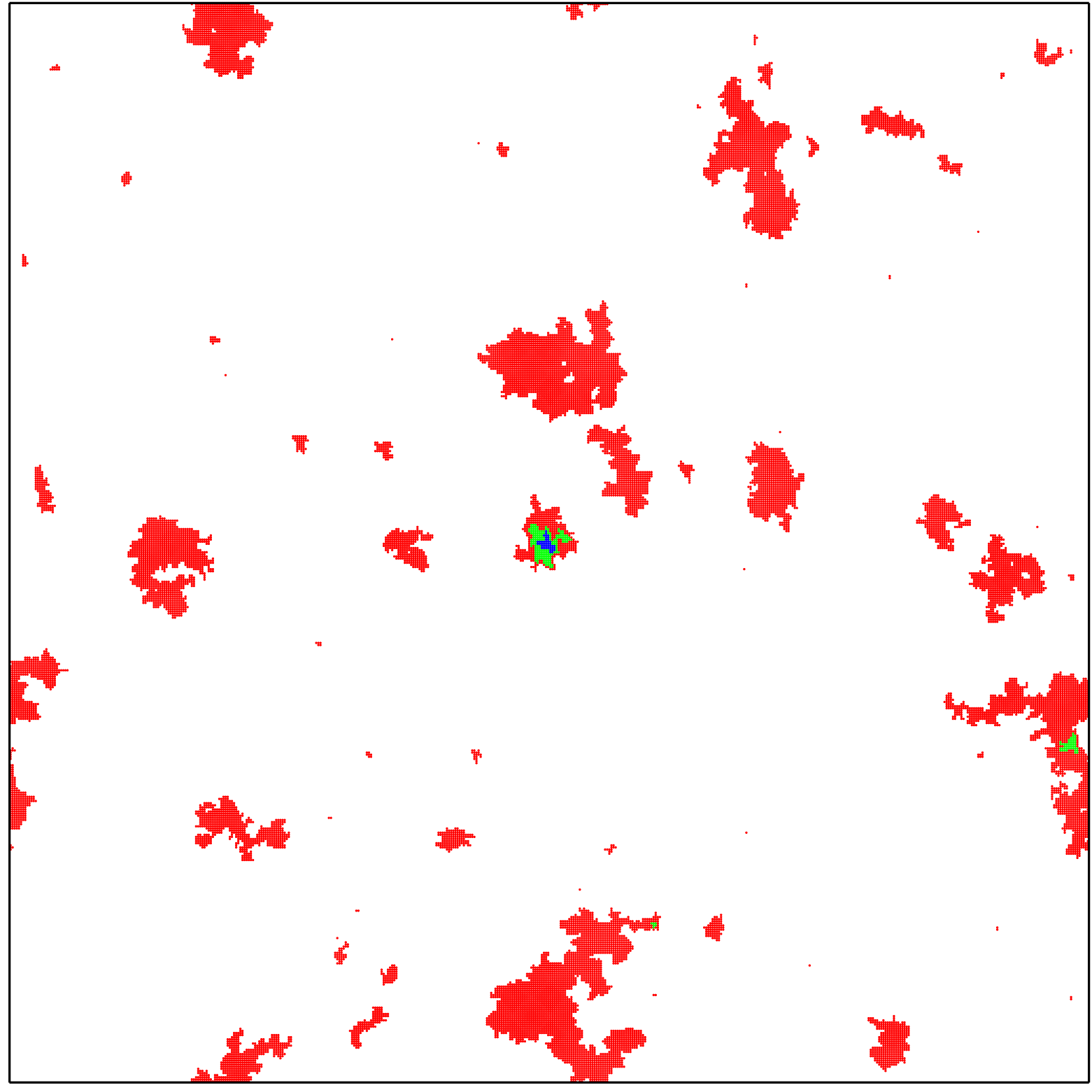,width=0.22\textwidth}\hfill}
\centerline{\hfill (c) $\phi=2^{-8}$, $(s<s_c)$ \hfill\hfill (d)
  $\phi=2^{-8}$, $(s>s_c)$\hfill}
\caption{\label{morph}(Color online) Morphology of avalanche clusters
  of DSM on SWN generated on $L=512$ square lattice for different
  values of $\phi$. (a) For $\phi=2^{-12}$, almost a BTW type
  cluster. (b) For $\phi=2^{-3}$, avalanche cluster on a RN completely
  scattered all over the lattice. (c) For $\phi=2^{-8}$, a small
  avalanche cluster of size ($352$) less than $s_c (\approx 360)$ is
  shown. Enlarged version of the same avalanche cluster is given in
  the inset. (d) For $\phi=2^{-8}$, a large avalanche cluster (a
  different realization than (c)) of size ($16872$) greater than
  $s_c$. Different colors correspond to different numbers of toppling
  of a node: red for $1$, green for $2$, blue for $3$, black for $4$,
  skyblue for $5$ and purple for more than $5$ toppling. No color
  corresponds to the nodes that did not topple at all during the
  avalanche. The black border represents the lattice boundary.}
\end{figure}

Morphology of avalanche clusters obtained in the steady state of DSM
on SWNs corresponding to different $\phi$ values are shown in
Fig.\ref{morph}. These avalanches are obtained on SWNs defined on a
square lattice of size $512\times 512$ for $\phi=2^{-12}$,
$\phi=2^{-8}$ and $\phi=2^{-3}$. Different colors correspond to
different numbers of toppling of a node. A typical avalanche cluster
in the RL regime with $\phi=2^{-12}$, is shown in
Fig.\ref{morph}(a). The avalanche cluster (of size $s=63774$) is
isotropic and mostly compact. It consists of concentric zones of lower
and lower number of toppling around the node with maximum number of
toppling (in purple) as expected for a BTW type avalanche cluster
\cite{grass,chris}. Few clusters of compact structure appear here and
there because of the presence of a small number of shortcuts in the
system a few sands grains are transported to the remote parts of the
lattice. However such a small distortion in the morphology of
avalanche cluster with respect to a single compact BTW type cluster is
not able to modify the scaling behavior.

In Fig.\ref{morph}(b), a typical avalanche cluster ($s=33567$)
obtained on RN corresponding to $\phi=2^{-3}$ is shown. In this case,
the avalanche cluster is completely scattered all over the network. A
large number of shortcuts are added to RL to make it RN and hence sand
grains from a toppled node of RN are transported to almost all other
nodes of the network through the shortcuts. The compact BTW type
cluster is therefore found scattered all over the lattice or the
network. Small patches of sites toppled only once are still
present. As $\phi$ approaches $1$, the size of such patches reduces.

The morphology of avalanche cluster of DSM on SWN with intermediate
$\phi$ is found either as that of BTW type cluster on RL or as sparse
clusters on RN. Two such avalanche clusters on SWN with $\phi=2^{-8}$
are shown in Fig.\ref{morph}(c) and (d). It is already seen that the
avalanche clusters on SWNs with intermediate $\phi$ follow two
different scaling forms below and above a characteristic toppling size
$s_c$. For $\phi=2^{-8}$, it is given by $s_c= \phi^{-1.06}\approx
360$. A cluster of size $352$ ($<s_c$) is shown in Fig.\ref{morph}(c)
and a cluster of size $16872$ ($>s_c$) is shown in
Fig.\ref{morph}(d). The larger cluster in Fig.\ref{morph}(d) is broken
into patches consisting of nodes mostly toppled once and scattered
over most of the network whereas the smaller cluster in
Fig.\ref{morph}(c) is still isotropic and compact. An enlarged version
of the small cluster is shown in the inset of
Fig.\ref{morph}(c). Therefore, on an SWN two types of clusters
appear. The smaller compact clusters ($s<s_c$) naturally follow BTW
type scaling and the large sparse clusters ($s>s_c$) follow MF type
scaling. The coexistence of two scaling forms on an SWN is then due to
the presence of both the clusters on the same network. As $\phi$
decreases (goes toward RL), $s_c$ becomes larger and consequently all
clusters are of BTW type. On the other hand as $\phi$ increases to $1$
(RN), $s_c$ goes down to $1$ and all the clusters are sparse and
scattered over all the nodes.

Sandpile dynamics then can be used as an useful tool to probe
different length scales present in the underlying structure on which
it is performed. The avalanches are expected to display appropriate
scaling behavior corresponding to different length scales.

\section{Distribution of area of various toppling sites}
\begin{figure}[t]
\begin{center}
\psfig{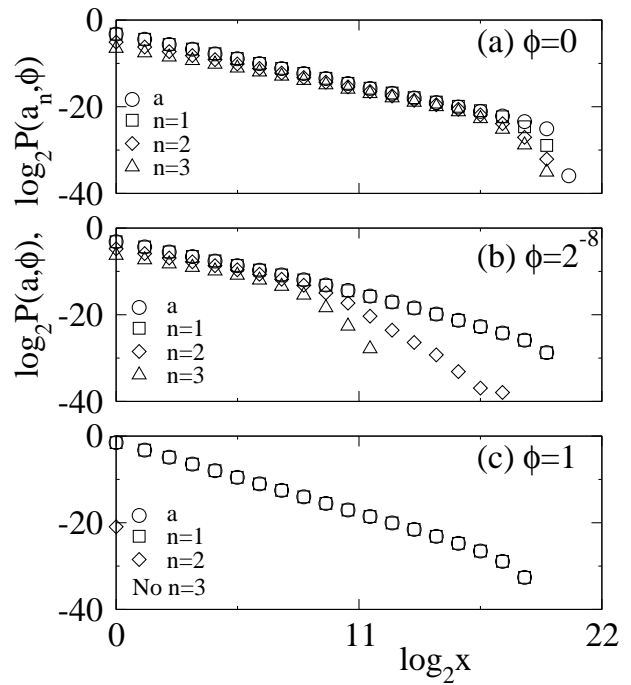}
\end{center}
\caption{Plot of $P(a,\phi)$ and $P(a_n,\phi)$ for $n=1,2,3$ in double
  logarithmic scale for different values of $\phi$: For $\phi=0$ (a),
  $\phi=2^{-8}$ (b) and $\phi=1$ (c) for the system size $L=1024$. }
\label{pnd3}
\end{figure}

From the morphologies of avalanche clusters, it is seen that on RL the
avalanches are consisting of sites toppled multiple times whereas on
RN they are mostly consisting of nodes toppled only once. On SWN with
intermediate $\phi$, clusters of both types appear. In order to
understand the type of sites present in an avalanche, the distribution
of area $a_n$ of sites those are toppled a fixed $n$ number of times
should be analyzed. Such area distributions for BTW model on RL were
found to obey power law scaling with exponents close to that of the
total area distribution exponent \cite{dhar_mann,banerjee}. The idea
of studying distributions of $a_n$ is extended here to the avalanches
obtained on SWNs. The scaling behavior of number of sites or nodes
that toppled $n$-times is then assumed to be
\begin{equation}
\label{pan}
P(a_n,\phi) \approx a_n^{-\tau_a^{(n)}(\phi)}
\end{equation}
where $n=1,2,3,\cdots$ corresponding to sites toppled only once, only
twice, only thrice, etc. In Fig.\ref{pnd3}, distribution functions
$P(a_n,\phi)$ corresponding to only once ($a_1$), only twice ($a_2$),
only thrice ($a_3$) are plotted and compared with the distribution of
total area $a$ for three different values of $\phi$, (a) $\phi=0$, (b)
$\phi=2^{-8}$ and (c) $\phi=1$ for $L=1024$. On RL, all three
distributions are extended over a long range of $a_n$, almost as large
as total area $a$. The distribution of $a_1$ has an exponent
$\tau_a^{(1)}=1.12\pm 0.01$ same as $\tau_a$. As the network grows to
an intermediate regime, say for $\phi=2^{-8}$, the distributions of
$a$ and $a_1$ are found to be almost same for all values of area as
given in Fig.\ref{pnd3}(b) whereas the distributions of $a_2$ and
$a_3$ are shrunk toward smaller areas. It can also be noticed that the
distribution of $a$ or $a_1$ has two different scaling forms
corresponding to two different regimes as it is
seen in the case of distribution of $s$ and $t$ (Fig.\ref{pds2}(a),
Fig.\ref{pds2}(b)). For $\phi=1$, the distribution of $a_1$ and that of
$a$ become inseparable as shown in Fig.\ref{pnd3}(c) and they have
same distribution exponent $\approx 1.5$ as that of the avalanche size
$s$ on a RN. This means that the avalanches are consisting of singly
toppled nodes and the difference between the avalanche area $a$ and
avalanche size $s$ disappears. The distribution of $a_2$ reduces to a
point and there is no node that toppled thrice or more. It can be
noted that $a_2=1$, {\em i.e.}; only one node has toppled twice. The
probability of occurrence of such an event is also very small,
$P(a_2,\phi=1)\approx 1/2^{20}$. It had already been noted that the
possibility of formation of loop in a branching process of toppling
events on a RN is vanishingly small and usually goes as $1/L^2$,
inverse of the number of nodes \cite{chris,ebon}. Thus the present
observation is consistent with the prediction of branching process.

\begin{figure}[t]
\begin{center}
\psfig{file=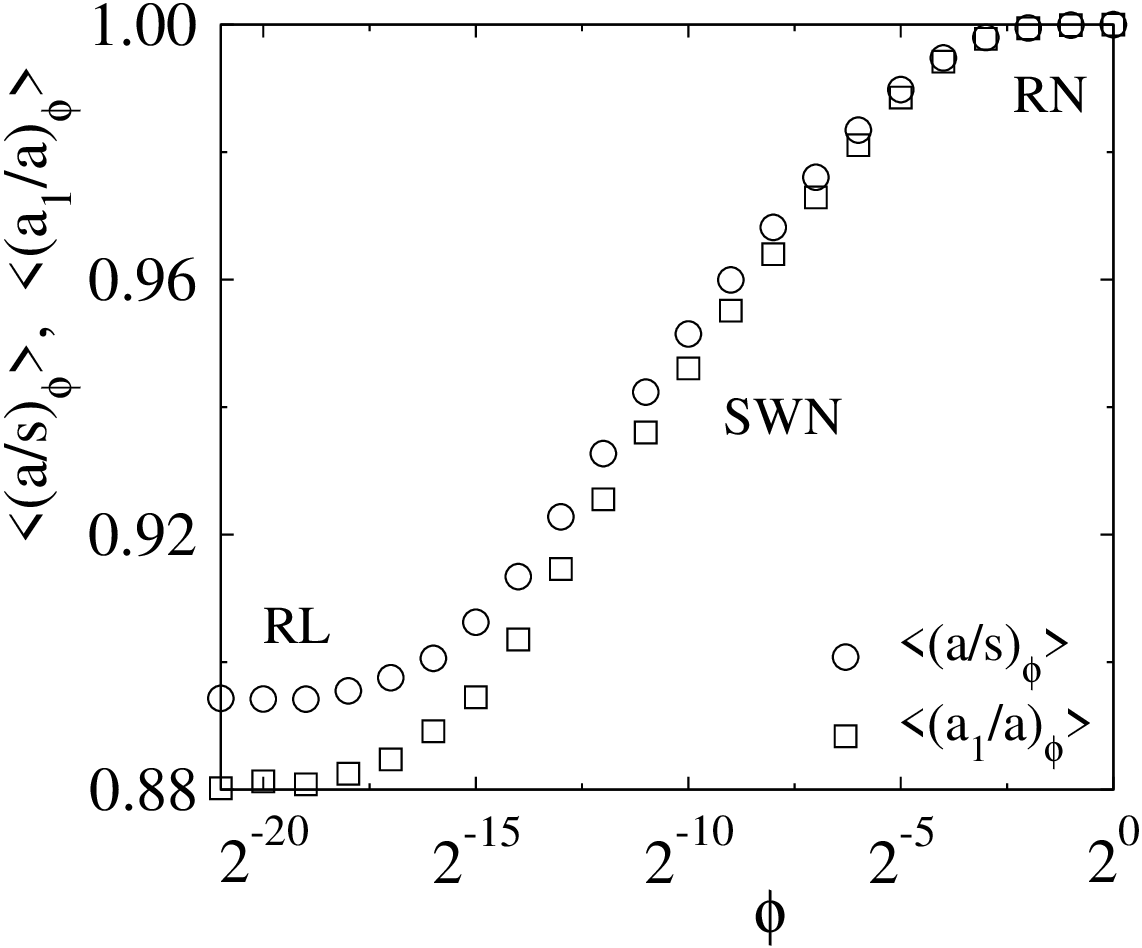,width=0.42\textwidth}
\end{center}
\caption {Plot of $\langle (a/s)_\phi\rangle$ ($\Circle$) and $\langle
  (a_1/a)_\phi\rangle$ ($\Box$) against $\phi$ for the system size $L=1024$.}
\label{sbya}
\end{figure}

Not only the probability distributions of $a$ and $s$ are same but
also the magnitude of $a$ and $s$ are found to be same on RN. This is
verified by calculating the ratio $\langle (a/s)_\phi\rangle$ and
$\langle (a_1/a)_\phi\rangle$ for several values of $\phi$. The
variation of $\langle (a/s)_\phi\rangle$ against $\phi$ is shown in
Fig.\ref{sbya} and compare with that of $\langle
(a_1/a)_\phi\rangle$. It can be seen that for $\phi\ge 0.1$ both the
ratios are one. They decrease as $\phi$ decreases. For $\phi < 0.1$,
the ratio $\langle (a/s)_\phi\rangle < 1$ indicates that $s>a$ and the
ratio $\langle (a_1/a)_\phi\rangle < 1$ indicates that $a>a_1$. It can
also be noted that $\langle (a/s)_\phi\rangle$ and $\langle
(a_1/a)_\phi\rangle$ are same for $\phi\ge 0.1$ whereas for $\phi<
0.1$ they are different.

\section{Time autocorrelation of toppling waves}
A toppling wave is the number of toppling during the propagation of
an avalanche starting from a critical site without toppling the same
site further. Each toppling of the critical site creates a new
toppling wave. The total number of toppling $s$ in an avalanche
can be considered as
\begin{equation}
\label{toppl}
s=\sum_{k=1}^m s_k
\end{equation}
where $s_k$ is the number of toppling in the $k$th wave and $m$ is the
number of toppling waves in an avalanche. The time evolution of
toppling dynamics then can be studied coarsening the avalanches into a
series of toppling waves \cite{prie}. The toppling waves generated in
the BTW model on RL were found to be correlated \cite{menech_corr}. As
a consequence of such correlation in the toppling waves, it was
observed that the model does not obey finite size scaling (FSS)
\cite{karmakar}. It is then interesting to study the time
autocorrelation of the toppling waves for DSM on SWNs to get a
limiting value of $\phi$ at which FSS would be obeyed for this
model. Following Menech and Stella \cite{menech_corr}, a time
autocorrelation function for an SWN with given $\phi$ is defined as
\begin{figure}[t]
\centerline{\hfill
\psfig{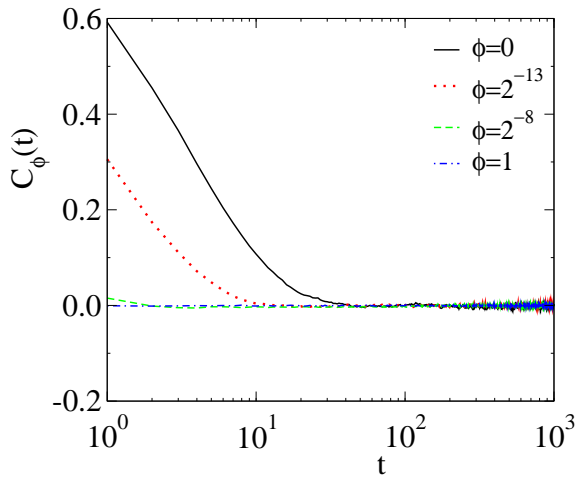} \hfill}
\caption{Plot of $C_\phi(t)$ against $t$ for different values of
  $\phi$: $\phi=0$ (in black solid line), $\phi=2^{-13}$ (in red
  dotted line), $\phi=2^{-8}$ (in green dashed line), and $\phi=1$ (in
  blue dashed dotted line) for $L=1024$.}
\label{tauto}
\end{figure} 
\begin{equation}
\label{corl}
C_\phi(t) = \frac{\langle s_{k+t}s_k\rangle -  \langle s_k\rangle^2}{
  \langle s_k^2\rangle - \langle s_k\rangle^2}
\end{equation}
where $t=1,2,\cdots$ and $\langle\cdots\rangle$ represents the time
average. $C_\phi(t)$ is calculated for four different values of
$\phi$, $\phi=0$, $\phi=2^{-13}$, $\phi=2^{-8}$ and $\phi=1$ on a
system of size $L=1024$, generating $2\times 10^{6}$ toppling waves in
the steady state for each $\phi$. $C_\phi(t)$s obtained for the above
$\phi$ values are plotted against $t$ in Fig.\ref{tauto}. It can be
seen that the toppling waves in DSM on the original lattice ($\phi=0$)
is positive (shown by a black solid line) and hence highly
correlated. Whereas on RN ($\phi=1$) it is always zero and hence
completely uncorrelated (shown by a blue dashed dotted line). As
$\phi$ increases from $0$ to $1$, the strength of positive correlation
decreases and vanishes at $\phi\approx 0.1$ corresponding to the onset
of RN. Zero autocorrelation in the toppling waves on RN is consistent
with the fact that the avalanches on such a network are consisting of
nodes mostly toppled only once. Since almost no node in an avalanche
toppled twice, an avalanche is then represented by a single toppling
wave. The toppling wave time series then consists of sequence of
toppling numbers of a single toppling wave of independent
avalanches. Hence, the toppling waves become uncorrelated. On the
other hand, the toppling waves of DSM on RL remain correlated as in
the case of BTW. It should be emphasized here that Karmakar {\em et al.}
\cite{karmakar} had shown that the toppling wave correlation in BTW
type sandpile model on a RL is essentially due to precise toppling
balance. Though in the present model on RN precise toppling balance is
present in the toppling rule, the toppling waves become
uncorrelated. Because on RN, the probability of formation of loop in
the toppling sequence is vanishingly small and hence the concept of
precise toppling balance become ineffective.

Since an avalanche cluster on RN consists of a single toppling wave,
the feedback to the original toppled node remains so low that in most
of the cases it never become upper critical again. Hence, in the
context of information propagation, the critical RN behaves like a
one way network. May be due to the fact that the RN already behaves
like a one way network, the sandpile on directed small world network
\cite{he} is found to belong to the same MF universality class.

\section{Diffusive to super diffusive sand transport}
The critical behavior of sandpile models on RL is believed to be
governed by the diffusive sand transport during avalanches. Since the
avalanche size (the total number of toppling) is equivalent to the
number of steps of a random walker starting from an arbitrary site to
reach the lattice boundary of a RL \cite{shilo,ssms}, their scaling
behavior with lattice size (or number of nodes) on SWNs is now
important to characterize. For a given $\phi$ and system size $L$, the
average number of steps $\langle n_\phi(L) \rangle$ required for a
random walker to reach the lattice boundary starting from an arbitrary
node of an SWN and the average avalanche size can be defined as
\begin{equation}
\label{nphi}
\langle n_\phi(L) \rangle =\int_0^{n_{max}} nP_\phi(n,L)dn  
\end{equation}
and
\begin{equation}
\label{sphi}
\langle s_\phi(L) \rangle =\int_0^{s_{max}} sP_\phi(s,L)ds
\end{equation}
where $P_\phi(n,L)$ is the probability to find a random walk with $n$
steps that reaches the lattice boundary starting from an arbitrary
node and $P_\phi(s,L)$ is the probability to have an avalanche of size
$s$ for the given $\phi$ and $L$. The scaling of $\langle n_\phi(L)
\rangle$ and $\langle s_\phi(L)\rangle$ with $L$ is assumed to be
\begin{figure}[t]
\begin{center}
\psfig{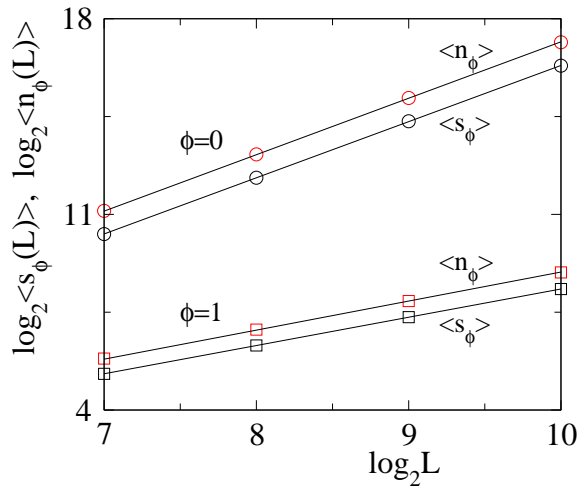}
\end{center}
\caption{(Color online) Average avalanche size $\langle
  s_\phi(L)\rangle$ (symbols in black) and $\langle n_\phi(L) \rangle$
  (symbols in red) is plotted against system size $L$ for $\phi=0$
  ($\Circle$) and $\phi=1$ ($\Box$). Solid line through the circles
  has slope $2$ and that through the squares has slope $1$.}
\label{av_s_t_vs_L}
\end{figure}
\begin{equation}
\label{nsL}
\langle n_\phi(L) \rangle \sim L^{\sigma_n(\phi)} \hspace{0.5cm}
  \mbox{and} \hspace{0.5cm} \langle s_\phi(L) \rangle = L^{\sigma_s(\phi)}
\end{equation}
where $\sigma_n(\phi)$ and $\sigma_s(\phi)$ are two exponents. In
order to verify such a scaling forms for $\langle n_\phi(L) \rangle$
and $\langle s_\phi(L) \rangle$, they are estimated as a function of
$L$ for $\phi=0$ and $\phi=1$. In Fig.\ref{av_s_t_vs_L}, $\langle
n_\phi(L) \rangle$ and $\langle s_\phi(L) \rangle$ are plotted against
$L$ in double logarithmic scale for both the values of $\phi$. The
values of the exponents $\sigma_n(\phi)$ and $\sigma_s(\phi)$ are
obtained as $\sigma_n = \sigma_s \approx 2$ for $\phi=0$ and $\sigma_n
= \sigma_s \approx 1$ for $\phi=1$. The solid lines are guide to eye
with respective slopes. Since at $\phi=0$, both $\langle n_\phi(L)
\rangle$ and $\langle s_\phi(L) \rangle$ scale as $\sim L^2$, the
random walk or the sand transport both are of diffusive nature whereas
at $\phi=1$ they scale as $\sim L$, therefore they are of super
diffusive nature. It is observed that the super diffusive nature
sustains over the whole RN region $\phi\ge 0.1$. But the diffusive
nature quickly dies out as the number of shortcuts increases in the
system. For intermediate values of $0<\phi<0.1$, no definite values of
$\sigma_n(\phi)$ or $\sigma_s(\phi)$ was possible to estimate because
the data did not represent a linear relationship on double logarithmic
scale. The curvature in the data is due to the fact that in the
intermediate region of $\phi$ both the scaling forms of $P(s,\phi)$
coexists. Therefore a crossover from diffusive to super diffusive sand
transport occurs as RL is evolved to RN. It can be seen that not only
the scaling of $\langle n_\phi \rangle$ and $\langle s_\phi \rangle$
are same but also the magnitude of $\langle n_\phi \rangle$ is just
twice of $\langle s_\phi \rangle$ for both $\phi=0$ and $\phi=1$ on a
given $L$. On RL it was already known that $\langle n_\phi \rangle =2
\langle s_\phi\rangle$ \cite{malcai,shilo}. Such a relationship is
then also valid on RN. It is also interesting to note that the
absolute values of $\langle s_\phi \rangle$ (or $\langle n_\phi
\rangle$) is much smaller on RN than on RL for a given $L$. It could
be recalled that the avalanches on RN consist of nodes that toppled
only once whereas on RL there exist sites that toppled multiple
times. The cut off of the distribution $P(s,\phi)$ on RN is much
smaller than that on RL (see Fig.\ref{pds}). Therefore, on RN
occurrence of an avalanche cluster with nodes toppled only ones,
consists of a single toppling wave, super diffusive sand transport
during an avalanche all are inter connected phenomena.

\begin{figure}[t]
\begin{center}
\psfig{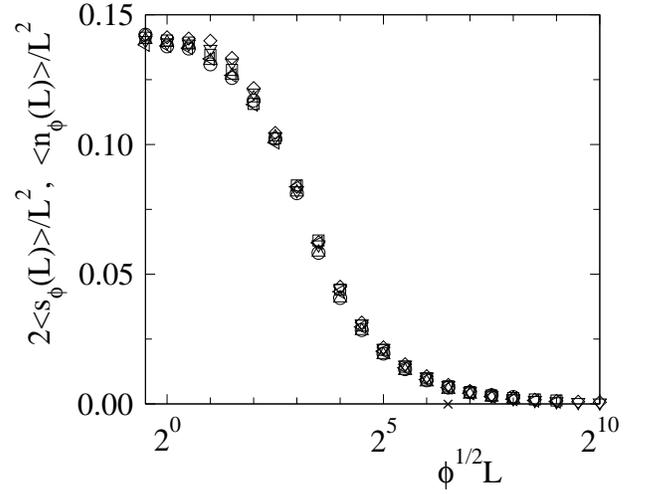} 
\end{center}
\caption{Plot of $2\langle s_\phi(L)\rangle/L^2$ and $\langle
  n_\phi(L)\rangle/L^2$ against the scaled variable
  $\phi^{1/2}L$. Different symbols for $\langle s_\phi(L)\rangle$ for
  different system size are: $\Circle$ for $L=256$, $\Box$ for $512$
  and $\Diamond$ for $1024$. For $\langle n_\phi(L)\rangle$ they are:
  $\triangle$ for $L=256$, $\triangleleft$ for $512$ and
  $\triangleright$ for $1024$. A reasonable data collapse is
  observed.}
\label{avg_s_coll}
\end{figure}

Since the shortest distance $\ell$ between two nodes of an SWN follows
two different scaling behavior given in Eq.\ref{l_scale1} and
\ref{l_scale2}, it would be interesting to verify whether $\langle
n_\phi(L) \rangle$ and $\langle s_\phi(L) \rangle$ follow a similar
scaling behavior on SWN or not. Following the scaling of $\langle \ell
\rangle$ given in Eq.\ref{l_scale2}, general scaling forms of
$\langle n_\phi(L) \rangle$ and $\langle s_\phi(L) \rangle$ are
proposed as
\begin{equation}
  \label{n_scale} 
  \langle n_\phi(L) \rangle = L^2\mathcal{G}(\phi^{1/2}L)
\end{equation}
and
\begin{equation}
  \label{s_scale} 
  \langle s_\phi(L) \rangle =\frac{1}{2} L^2 \mathcal{G}(\phi^{1/2}L)
\end{equation} 
where $\mathcal{G}(x)$ is a universal scaling function given by
\begin{eqnarray}
\label{gscaling3}
 \mathcal{G}(x)&\propto& \left \{ \begin{array}{ll}
  \mbox{constant,} & x \ll 1
  \\1/x,  & x \gg 1 \\
       \end{array} \right. 
\end{eqnarray}
Verification of the above scaling form is performed by estimating
$\langle n_\phi(L) \rangle$ and $\langle s_\phi(L) \rangle$ for
different $L$ for the whole range of $\phi$ between $0$ and $1$. In
Fig.\ref{avg_s_coll}, $\langle n_\phi(L)/L^2 \rangle$ and $2\langle
s_\phi(L) \rangle/L^2$ are plotted against the scaled variable
$x=\phi^{1/2}L$.  Reasonable data collapses are observed for both
$\langle n_\phi(L) \rangle$ and $\langle s_\phi(L) \rangle$. It should
be noted here that on a two dimensional regular square lattice
$\langle n_\phi \rangle \approx aL+bL^2$, where $a=0.56$ and $b=0.14$
for small $L$ \cite{shilo}. However in the limit $L\rightarrow
\infty$, such a scaling can be approximated as $\langle n_\phi(L)
\rangle \approx 0.14 L^2$. In the limit $\phi\rightarrow 0$, the
scaling function approaches $0.14$. On the other hand, the $1/x$
scaling would be valid on RN, i.e. for $\phi\ge 0.1$. For the lowest
lattice size the corresponding value of the scaled variable is marked
by a cross on the horizontal axis beyond which $1/x$ scaling is
expected to be valid. It should also be noted here that the number of
distinct nodes $S(n)$ visited by a random walker in $n$ time steps on
a $1d$ SWN for a fixed $L$ and $\phi$ represents a crossover in
scaling from $S(n)\sim \sqrt{n}$ for $n\ll \xi^2$ to $S(n)\propto n$
for $n\gg \xi^2$ \cite{jasch,lahtinen,almaas}.

\section{FSS on random network}
Since SWNs are generated on finite systems of size $L$ with a given
$\phi$, the probability distributions of avalanche quantities then
should depend on $L$. The scaling form of the distribution function is
assumed to be
\begin{equation}
\label{pdfss}
P_\phi(x,L) = x^{-\tau_x(\phi)} f[x/L^{D_x(\phi)}]
\end{equation}
where $x\in \{s,a,t\}$ and $D_x(\phi)$ is the capacity dimension of
the avalanche property $x$ on SWN with a given $\phi$.  It was
observed that the avalanche properties like $s$ or $t$ of BTW type
models do not follow FSS ansatz on RL \cite{menech_moment,lubeck}. In
the following, the FSS analysis is performed for $s$ and $t$ on RN as
well as on SWNs employing moment analysis
\cite{menech_moment,lubeck,karmakar}. The average $q$th moment of an
avalanche property $x$ for a given $\phi$ can be obtained as
\begin{eqnarray}
\label{qmnts}
\langle x_\phi^q(L) \rangle &=& \int_0^{x_{max}}x^qP_\phi(x,L)dx\nonumber\\ 
&=&\int_0^{x_{max}}x^{q-\tau(\phi)}f[x/L^{D_x(\phi)}]dx.
\end{eqnarray}
Hence, the system size dependence of $\langle s_\phi^q(L) \rangle$ and
$\langle t_\phi^q(L) \rangle$ are expected to be
\begin{equation}
\label{sqphi}
\langle s_\phi^q(L) \rangle \sim
L^{\sigma_s(q,\phi)}\hspace{0.5cm}\mbox{and}\hspace{0.5cm}\langle
t_\phi^q(L) \rangle \sim L^{\sigma_t(q,\phi)}
\end{equation}
where,
\begin{equation}
\label{sigphi}
 \sigma_x(q,\phi)=[q+1-\tau_x(\phi)]D_x(\phi).
\end{equation}
for $x\in{\{s,t\}}$ and $q=1$ corresponds to the average values of the
respective avalanche properties such as $\langle s_\phi \rangle$,
$\langle t_\phi \rangle$, etc. For $P_\phi(x,L)$ to obey FSS for a
given $\phi$ value, the moment exponent $\sigma_x(q,\phi)$ should have
a constant gap between two successive values of $q$, {\em i.e.};
$\sigma_x(q+1,\phi) - \sigma_x(q,\phi) = D_x(\phi)$ for the respective
$\phi$ value. For avalanche properties it was usually found that the
gap converge to the respective capacity dimension as $q\rightarrow
\infty$. In order to determine $D_s(\phi)$ and $D_t(\phi)$, sequences
of exponents $\sigma_s(q,\phi)$ and $\sigma_t(q,\phi)$ are obtained
for $400$ equally spaced values of $q$ between $0$ and $4$ for several
$\phi$ values. The constant gap between two successive
$\sigma_x(q,\phi)$s is then verified by estimating the slope
$\partial\sigma_x(q,\phi)/\partial q$ using finite difference method.

\begin{figure}[t]
\begin{center}
\psfig{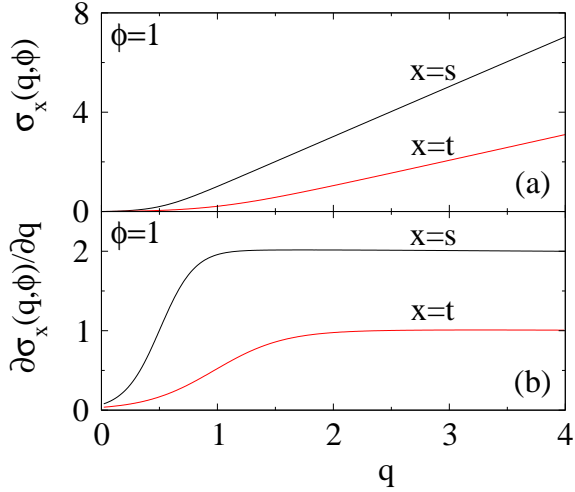}
\end{center}
\caption{(a) $\sigma_x(q,\phi)$ with $x\in\{s,t\}$ is plotted against
  $q$ for $\phi=1$. (b) $\partial\sigma_x(q,\phi)/\partial q$ is
  plotted against $q$ for $\phi=1$. It converges to $2$ for $x=s$ and
  to $1$ for $x=t$.}
\label{sigma_s}
\end{figure}

For $\phi<0.1$, the finite differences $\partial\sigma_s(q,\phi)/\partial
q$ and $\partial\sigma_t(q,\phi)/\partial q$ for both the sequences of
$\sigma_s(q,\phi)$ and $\sigma_t(q,\phi)$ did not converge to any
finite value upto $q=4$. Hence, $P_\phi(x,L)$, $x\in\{s,t\}$ does not
follow FSS ansatz in the SWN regime. Note that on SWN both the scaling
coexist. Since $P_\phi(x,L)$ for $x<x_c$ does not follow FSS, the
distribution functions for the full range of $x$ is then expected not
to follow FSS.   

For $0.1 \le \phi\le 1$, FSS is expected to be valid and it is
verified for several values of $\phi$ in this region. Data for
$\phi=1$ is presented below. The variation of $\sigma_s(q,\phi)$ and
$\sigma_t(q,\phi)$ for $\phi=1$ are plotted against moment $q$ in
Fig.\ref{sigma_s}(a) and that of $\partial\sigma_s(q,\phi)/\partial q$
and $\partial\sigma_t(q,\phi)/\partial q$ against moment $q$ are shown
in Fig.\ref{sigma_s}(b). It can be seen that for $\phi=1$ the
derivatives saturate to $D_s \approx 2$ and $D_t \approx 1$ for higher
values of $q$. The value of $D_s$ for $\phi=1$ is expected to be $2$
because on RN all avalanches are constituted of sites toppled only
once, that is to say the avalanche area and avalanche size have no
difference. On the other hand, the value of $D_t$ for $\phi=1$ is
expected to be $1$ because on RN all avalanches are constituted of
single toppling wave, that is to say the number of parallel updates in
a single toppling wave is proportional to the system size $L$ (see
Fig.\ref{pds}(b)). On the RL it was known that
$D_s/D_t=\gamma_{st}$. Such a scaling relation is also valid on
RN. Since $D_s= 2$ and $D_t=1$ for $\phi=1$, the value of
$\gamma_{st}$ is expected to be two as it is estimated in section
V. For $q=1$ and $\phi=1$, the scaling relations
$\sigma_s=(2-\tau_s)D_s$ and $\sigma_t=(2-\tau_t)D_t$ are expected to
be satisfied. Since $\tau_s=3/2$ and $D_s=2$ for $\phi=1$, the value
of $\sigma_s$ must be one as measured on RN. Similarly, from the other
scaling relation $\sigma_t$ for $q=1$ is expected to be zero because
$\tau_t=2$.  Numerically a small finite value of $\sigma_t$ for $q=1$
is estimated. However, for $q=2$, $\sigma_t=1$ as expected.

\begin{figure}[t!]
\begin{center}
\psfig{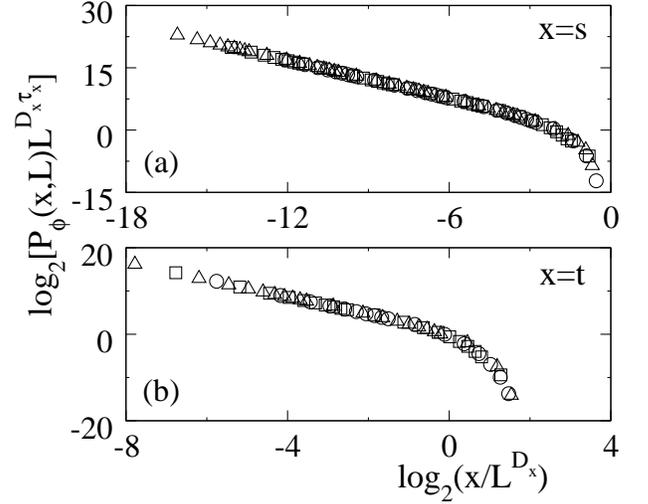}
\end{center}
\caption{Plot of scaled distribution $P_\phi(x,L)L^{D_x\tau_x}$
  against the scaled variable $x/L^{D_x}$ for $\phi=1$ in (a) for
  $x=s$ and in (b) for $x=t$. Different symbol correspond to different
  system size as $L=256(\Circle)$, $512(\Box)$ and
  $1024(\triangle)$. A reasonable data collapse is obtained for both
  $s$ and $t$.}
\label{col_s}
\end{figure}

Finally the scaling function forms of $P_\phi(s,L)$ and $P_\phi(t,L)$
for $\phi=1$ on RN are verified by data collapse. In
Fig.\ref{col_s}(a), the scaled probability distribution
$P_\phi(s,L)L^{D_s\tau_s}$ for $\phi=1$ is plotting against the scaled
variable $s/L^{D_s}$ taking $\tau_s=3/2$ and $D_s=2$ as for RN. In
Fig.\ref{col_s}(b), $P_\phi(t,L)L^{D_t\tau_t}$ for $\phi=1$ is plotting
against the scaled variable $t/L^{D_t}$ taking $\tau_t=2$ and $D_t=1$
as for RN. It can be seen that a reasonable data collapse is obtained
for RN generated on three different systems sizes $L=256, 512$ and
$1024$. The assumed FSS form of $P_\phi(x,L)$ on the RN is then
rightly chosen. Therefore, FSS would be valid even for a sandpile
model with deterministic and conservative toppling rules along with
complete toppling balance if it is defined on a system in which
concept of neighborhood does not exist and has long distance
connectivity.

\section{Conclusion}
A generalized DSM is constructed and studied on SWNs. Apart from BTW
type correlated scaling for $\phi \lesssim 2^{-12}$, two important
characteristic features of DSM, one on SWN for $2^{-12} <\phi<0.1$ and
the other on RN for $0.1 \le\phi\le 1$ are identified and
characterized. First, DSM on SWNs exhibits two scaling behaviors
simultaneously. One is that of BTW type scaling on RL and the other is
that of MF scaling on RN corresponding to existence of strong
neighborhood as that of RL as well as vanishingly small shortest
distance between two nodes as that of RN on an SWN. A characteristic
value of every avalanche property was possible to identify around
which coexistence scaling of probability distribution functions are
proposed and numerically verified. The avalanche clusters following
BTW scaling are compact BTW type clusters whereas those following MF
scaling are sparse and scattered all over the network. Since avalanche
clusters segregate as per the length scales of the SWN, sandpile
dynamics can be used as a probe to identify different length scales
present in the underlying structure on which it is performed. Second,
FSS is found to be valid for DSM on RN in contrary to the fact that
DSM does not follow FSS on RL or on SWN. The validity of FSS on RN for
DSM is due to the fact that the avalanches on RN are consisting of
nodes toppled only once. The probability of appearance of a node that
toppled more than once is vanishingly small on RN as the number of
nodes $N\rightarrow \infty$. As a consequence, precise toppling
balance becomes ineffective as well as toppling waves become
uncorrelated. Because of the presence of long distance connections,
sand transport becomes super diffusive on RN though it was diffusive
on RL. Super diffusive sand transport is found to be essential in
order to satisfy finite size scaling relations. Therefore, BTW type
correlated sandpile models would also follow FSS if they are studied
on systems without spatial structure and have long distance
connections.

\section{Acknowledgments}
Financial assistance from DST, Government of India through project
No. SR/S2/CMP-61/2008 is gratefully acknowledged.

\end{document}